\documentclass[useAMS,usenatbib]{mn2e}
\onecolumn
\usepackage{color}
\usepackage{verbatim}

\newcommand{\dfrac}{\frac}
\title[SPE equations of motion]{The equations of motion of a secularly precessing elliptical orbit}
\author[S. Casotto and M. Bardella]{S. Casotto$^{1,2}$\thanks{E-mail:
stefano.casotto@unipd.it} M. Bardella$^{2}$\\
$^{1}$Dipartimento di Fisica e Astronomia, Universit\`{a} di Padova, Vicolo dell'Osservatorio 3, 35122 Padova, Italy  \\
$^{2}$Centro Interdipartimentale Studi e Attivit{\`a} Spaziali, Universit\`{a} di Padova, Via Venezia 15, 35131 Padova, Italy}
\begin{document}

\date{In original form 2012 August 10}

\pagerange{\pageref{firstpage}--\pageref{lastpage}} \pubyear{2012}

\maketitle

\label{firstpage}

\begin{abstract}
The equations of motion of a secularly precessing ellipse are developed
using time as the independent variable. The equations are useful when
integrating numerically the perturbations about a reference trajectory which
is subject to secular perturbations in the node, the argument of pericenter
and the mean motion. Usually this is done in connection with Encke's method
to ensure minimal rectification frequency. Similar equations are already
available in the literature, but they are either given based on the true
anomaly as the independent variable \citep{KynerBennett:1966}, or in mixed mode
with respect to the time through the use of a supporting equation to track
the anomaly \citep{Escobal1965}. The equations developed here form a complete and
independent set of six equations in the time. Reformulations both of
Escobal's and Kyner and Bennett's equations are also provided which
lead to a more concise form. 

\end{abstract}


\section{Introduction}

In many cases one is confronted with the evaluation of the spectral
characteristics of specific perturbation accelerations on the orbital motion
of planetary satellites. In several instances analytical evaluations can be
provided. However, this is not always possible, and occasionally one may
desire verification of the analytical results. Since the analytical
development of perturbations is always based on some reference orbit, it
becomes of interest in numerical verification work to be able to reproduce
the same reference orbit. The simplest such orbit is an invariable ellipse,
and Encke's method can then be adopted for the numerical integration of
perturbations about this nominal orbit. Encke's method can be generalized to
include non-Keplerian reference trajectories. The method can then be
formulated in terms of the differential equation of the reference orbit $%
\bmath{r}^{\ast }\left( t\right) $ and the equation of the perturbation
of the relative position $\delta \bmath{r}=\bmath{r}\left(
t\right) -\bmath{r}{^{\ast }}\left( t\right) $ of the trajectory 
$\bmath{r}\left( t\right) .$ These equations are
\begin{eqnarray}
\frac{d^{2}\bmath{r}^{\ast }}{dt^{2}} &=&-\,\frac{\mu }{r^{\ast 3}}%
\bmath{r}^{\ast }+\bmath{s}\left( \bmath{r}^{\ast },%
\bmath{\dot{r}}^{\ast },t\right) ,  \label{Encke EoM 1} \\
\frac{d^{2}\delta \bmath{r}}{dt^{2}} &=&-\frac{\mu }{r^{\ast 3}}\left\{
f\left( q\right) \bmath{r}^{\ast }+\left[ \,1+f\left( q\right) \right]
\delta \bmath{r}\right\} +\bmath{g}\left( \bmath{r},%
\bmath{\dot{r}},\bmath{r}^{\ast },\bmath{\dot{r}}^{\ast
},t\right) ,  \label{Encke EoM 2}
\end{eqnarray}%
where $\mu =G\left( m_{1}+m_{2}\right) $ is the gravitational parameter of
two masses $m_{1}$ and $m_{2},$ $\bmath{s}\left( \bmath{r},%
\bmath{\dot{r}},t\right) $ is the disturbing acceleration which
generates the non-Keplerian nominal trajectory, while $\bmath{g}$ is
the \emph{complementary} disturbing acceleration. This is defined as the
difference between the actual disturbing acceleration $\bmath{p}\left( 
\bmath{r},\bmath{\dot{r}},t\right) $ acting on the body whose
orbit is being propagated and the disturbing acceleration of the
nominal trajectory,  
\begin{equation}
\bmath{g}\left( \bmath{r},\bmath{\dot{r}},\bmath{r}%
^{\ast },\bmath{\dot{r}}^{\ast },t\right) =\bmath{p}\left( 
\bmath{r},\bmath{\dot{r}},t\right) -\bmath{s}\left( 
\bmath{r}^{\ast },\bmath{\dot{r}}^{\ast },t\right) .
\label{complementary acceleration}
\end{equation}
Encke's parameter $q$ is defined \citep{Battin1987} as%
\begin{equation}
q=-\frac{\left( 2\bmath{r}^{\ast }+\delta \bmath{r}\right) \cdot
\delta \bmath{r}}{\left( \bmath{r}^{\ast }+\delta \bmath{r}%
\right) \cdot \left( \bmath{r}^{\ast }+\delta \bmath{r}\right) },
\label{Encke q non singular}
\end{equation}%
and the auxiliary function $f\left( q\right) $ has the form%
\begin{equation}
f\left( q\right) =q\frac{3\left( 1+q\right) +q^{2}}{1+\left( 1+q\right)
^{3/2}}.  \label{Encke f(q) non singular}
\end{equation}%
Note that in the unperturbed case $(\bmath{s} = \bmath{p}\equiv 
\bmath{0})$ Encke's equations (\ref{Encke EoM 1}) and (\ref{Encke EoM 2})
are exactly equivalent to the Two-Body equations of motion of the two
trajectories. Of course, in this case, their use is limited to
non-osculating initial conditions. In general, however, the initial
conditions for the perturbed case are such that $\delta \bmath{r}\left(
t_{0}\right) =\delta \bmath{\dot{r}}\left( t_{0}\right) =\bmath{0}.$
Also note that Encke's equations are completely equivalent to the original
perturbation problem stated for trajectories $\bmath{r}^{\ast }\left(
t\right) $ and $\bmath{r}\left( t\right) .$ In analytical developments
the approximation is usually made, however, of evaluating the disturbing
acceleration $\bmath{p}$ on the nominal trajectory, that is, of
replacing $\bmath{g}\left( \bmath{r},\bmath{\dot{r}},%
\bmath{r}^{\ast },\bmath{\dot{r}}^{\ast },t\right) $ with $%
\bmath{g}\left( \bmath{r}^{\ast },\bmath{\dot{r}}^{\ast
},t\right) =\bmath{p}\left( \bmath{r}^{\ast },\bmath{\dot{r}}%
^{\ast },t\right) -\bmath{s}\left( \bmath{r}^{\ast },\bmath{%
\dot{r}}^{\ast },t\right) .$ 

\cite{KynerBennett:1966} (hereafter K\&B) proposed a modification to Encke's method
which replaces the invariable ellipse with a precessing ellipse that
includes the first-order secular effects of the Earth's oblateness. This has
the desirable effect of limiting the frequency of rectifications necessary
to contain within specified bounds the unavoidable divergence of the
perturbed orbit from the reference orbit. The K\&B formulation is given
using the true anomaly as the independent variable. For comparison with
analytical theories based on time as the independent variable, like Kaula's
linear satellite theory \citep{Kaula}, the K\&B approach needs to be reformulated. This
has been done by \cite{Escobal1965}. More recently \cite{Lundberg2000}, on the basis of 
previous work by \cite{Lundberg1991}, have introduced an extension of Encke's 
method for application to long-arc orbit determination which uses a precessing and librating ellipse 
of variable shape based on the Escobal model. The behaviour of the  K\&B and the Escobal reference orbits
are similar, but the K\&B approach contains some periodic terms in addition to the purely secular terms of
the time-wise approach of Escobal. Comparison with Kaula's
theory requires a time-wise approach, but implementation of Escobal's
original equations is cumbersome since they require the use of a supporting
equation for the true anomaly, also a feature of the K\&B approach. It is then
desirable to investigate the possibility of improving on the available
representations of the secularly precessing ellipse. In the following we will review these classical methods and
provide for each an alternative formulation. Finally we will develop a novel formulation with the time as 
independent variable, but distinctly different from Escobal's, which leads
to a vector differential equation of the second order, whose coefficients
are functions of the dynamical state variables only.

\section{The Secularly Precessing Ellipse}\label{SPE}

The secularly precessing ellipse (SPE) can be defined through the secular
rates of the angular elements either as a function of the true anomaly or as
a function of the mean anomaly. Here we define the SPE for both cases
providing analytical expressions for the secular rates of the angular
elements due to the first zonal harmonic of the gravity field of the central
body.

\subsection{True anomaly $f$ as independent variable}\label{SPE_true}

Given the gravitational parameter $\mu _{0}=G\left( m_{1}+m_{2}\right) ,$
let the orbital elements at epoch be $(a_{0},e_{0},i_{0},\Omega
_{0},\omega _{0},M_{0})$ where the associated Keplerian mean motion is $%
n_{0}=\left( \mu _{0}/a_{0}^{3}\right) ^{1/2}.$ In the presence of a second
degree zonal harmonic potential $J_{2}=-C_{20},$ the secular variations of
the elements can be given as a function of the true anomay $f$ as 
\begin{eqnarray}
a &=&a_{0},\qquad \Omega \left( f\right) =\tau \left( f-f_{0}\right) +\Omega
_{0},  \nonumber \\
e &=&e_{0},\qquad \omega \left( f\right) =\eta \left( f-f_{0}\right) +\omega
_{0},  \label{TA_elorb} \\
i &=&i_{0},\qquad M\left( t\right) =n\left( t-t_{0}\right) +M_{0},  \nonumber
\end{eqnarray}%
where the perturbed secular mean motion is given by 
\begin{equation}
n=n_{0}\left( 1-\gamma \right) ,  \label{n}
\end{equation}%
and the constant secular rates $\tau $ and $\eta $ due to the first zonal
harmonic are given by \cite{Sterne1960}, \cite{KynerBennett:1966} 
\begin{eqnarray}
\tau  &=&\frac{d\Omega }{df}=-\frac{3}{2}\frac{J_{2}}{\left( 1-e^{2}\right)
^{2}}\left( \frac{a_{e}}{a}\right) ^{2}\cos i,  \label{TA_rates tau} \\
\eta  &=&\frac{d\omega }{df}=\frac{3}{4}\frac{J_{2}}{\left( 1-e^{2}\right)
^{2}}\left( \frac{a_{e}}{a}\right) ^{2}\left( 4-5\sin ^{2}i\right) ,
\label{TA_rates eta}
\end{eqnarray}%
with%
\begin{equation}
\gamma =-\frac{3}{2}J_{2}\left( \frac{a_{e}}{a}\right) ^{2}\left( \frac{a}{%
r_{0}}\right) ^{3}\left[ 1-3\sin ^{2}i\sin ^{2}\left( \omega
_{0}+f_{0}\right) \right] .  \label{TA gamma}
\end{equation}%
where $r_{0}=a\left( 1-e^{2}\right) /\left( 1+e\cos f_{0}\right) .$ Note
that the phoronomic elements $a,$ $e$ and $i$ do not show secular behavior.

\subsection{Mean anomaly $M$ as independent variable}\label{SPE_mean}

If we choose time $t$ as the independent variable, then the secular rates of
the orbital elements assume the expressions 
\begin{eqnarray}
a &=&a_{0}, \qquad \Omega \left( t\right) =\dot{\Omega}\left( t-t_{0}\right)
+\Omega _{0},  \nonumber \\
e &=&e_{0}, \qquad \omega \left( t\right) =\dot{\omega}\left( t-t_{0}\right)
+\omega _{0},  \label{MA_elorb} \\
i &=&i_{0}, \qquad M\left( t\right) =\bar{n}\left( t-t_{0}\right) +M_{0}, 
\nonumber
\end{eqnarray}%
where the constant secular rates $\dot{\Omega}$ and $\dot{\omega}$ due to
the first zonal harmonic are given by \cite{Kaula} 
\begin{eqnarray}
\dot{\Omega} &=&\frac{3}{2}\frac{n_{0}C_{20}}{\left( 1-e^{2}\right) ^{2}}%
\left( \frac{a_{e}}{a}\right) ^{2}\cos i,  \label{MA_rates N} \\
\dot{\omega} &=&\frac{3}{4}\frac{n_{0}C_{20}}{\left( 1-e^{2}\right) ^{2}}%
\left( \frac{a_{e}}{a}\right) ^{2}\left( 1-5\cos ^{2}i\right) ,
\label{MA_rates omega}
\end{eqnarray}%
and the perturbed mean motion $\bar{n}$ is 
\begin{equation}
\bar{n}=n_{0}\left( 1-\bar{\gamma}\right) ,  \label{n_bar}
\end{equation}%
with%
\begin{equation}
\bar{\gamma}=-\frac{3}{4}\frac{C_{20}}{\left( 1-e^{2}\right) ^{3/2}}\left( 
\frac{a_{e}}{a}\right) ^{2}\left( 1-3\cos ^{2}i\right) .  \label{MA gamma}
\end{equation}

\subsection{Mapping the SPE to cartesian state}\label{SPE_mapping}

Equations (\ref{TA_elorb}) and equations (\ref{MA_elorb}) provide at any
time $t$ the secular evolution of the orbit elements of the reference orbit.
As noted above, the two reference orbits are not exactly the same since the
true anomaly is a periodic function of time. Given the initial Keplerian
state vector $(a_{0},e_{0},i_{0},\Omega _{0},\omega _{0},M_{0})$ at epoch $%
t_{0},$ and the secular rates (\ref{TA_rates tau}) and (\ref{TA_rates eta})
and the constant (\ref{TA gamma}) in the case of true anomaly as independent
variable, or the secular rates (\ref{MA_rates N}) and (\ref{MA_rates omega})
and the constant (\ref{MA gamma}) if the mean anomaly is adopted as the
independent variable, the nominal or reference trajectory $\bmath{r}%
^{\ast }(t)$---hereafter simply indicated with $\bmath{r}(t)$ for
conciseness---is completely specified at any time $t.$\textbf{\ }The
corresponding Cartesian state $\bmath{r}(t),\bmath{v}(t)$ can be
computed by the following procedure

\begin{enumerate}
\item Compute the nominal true anomaly $f$ by solving first the modified
Kepler's equation%
\begin{equation}
\tilde{n}\left( t-t_{0}\right) +M_{0}=E-e_{0}\sin E,  \label{Kepler_mod}
\end{equation}%
for the eccentric anomaly $E$ and then by transforming to $f$ via the usual
Gauss formula%
\begin{equation}
\tan \frac{f}{2}=\sqrt{\frac{1+e_{0}}{1-e_{0}}}\tan \frac{E}{2}.
\end{equation}%
The modified mean motion $\tilde{n}$ is either $n$ from equation (\ref{n})
or $\bar{n}$ from equation (\ref{n_bar}), depending respectively on whether
the true anomaly or the time is assumed as the independent variable.

\item Compute the radius vector $r$ and the radial velocity $v_{r}$ by means
of 
\begin{equation}
r=\frac{a\left( 1-e^{2}\right) }{1+e\cos f},\qquad v_{r}=\dot{r}=\frac{%
\tilde{n}ae\sin f}{\sqrt{1-e^{2}}}.  \label{pos_vel}
\end{equation}

\item Finally position $\bmath{r}(t)$ and velocity $\bmath{v}(t)$
are given by%
\begin{eqnarray}
\bmath{r}(t) &=&\mathbfss{R}\left( \Omega ,i,\omega +f\right) \bmath{%
r}_{rtn},  \label{nominal_pos} \\
\bmath{v}(t) &=&{\dot\mathbfss{R}}\left( \Omega ,i,\omega +f\right) 
\bmath{r}_{rtn}+\mathbfss{R}\left( \Omega ,i,\omega +f\right) \bmath{%
\dot{r}}_{rtn},  \label{nominal_vel}
\end{eqnarray}%
where%
\begin{equation}
\bmath{r}_{rtn}=\left( 
\begin{array}{c}
r \\ 
0 \\ 
0%
\end{array}%
\right) ,\hspace{0.9cm}\bmath{\dot{r}}_{rtn}=\left( 
\begin{array}{c}
\dot{r} \\ 
0 \\ 
0%
\end{array}%
\right) ,\hspace{0.9cm}\bmath{\ddot{r}}_{rtn}=\left( 
\begin{array}{c}
\ddot{r} \\ 
0 \\ 
0%
\end{array}%
\right) ,
\end{equation}%
the vector $\bmath{\ddot{r}}_{rtn}$ having been defined for later use,
and the rotation matrix $\mathbfss{R}\left( \Omega ,i,\omega +f\right) $ from
the orbital reference frame (the RTN frame, defined by the local radial, 
transverse, and normal directions) to the inertial reference frame is given by 
\begin{equation}
\mathbfss{R}\left( \Omega ,i,\omega +f\right) =\mathbfss{D}\left( \Omega \right) 
\mathbfss{C}\left( i\right) \mathbfss{B}\left( \omega +f\right) ,  \label{R_Mat}
\end{equation}%
with the elementary rotations defined as%
\begin{equation}
\mathbfss{D}\left( \Omega \right)  =\left( 
\begin{array}{ccc}
\cos \Omega  & -\sin \Omega  & 0 \\ 
\sin \Omega  & \cos \Omega  & 0 \\ 
0 & 0 & 1%
\end{array}%
\right) ,  \label{D_Mat} 
\end{equation}
\begin{equation}
\mathbfss{C}\left( i\right)  =\left( 
\begin{array}{ccc}
1 & 0 & 0 \\ 
0 & \cos i & -\sin i \\ 
0 & \sin i & \cos i%
\end{array}%
\right) ,  \label{C_Mat} 
\end{equation}
\begin{equation}
\mathbfss{B}\left( \omega +f\right)  =\left( 
\begin{array}{ccc}
\cos \left( \omega +f\right)  & -\sin \left( \omega +f\right)  & 0 \\ 
\sin \left( \omega +f\right)  & \cos \left( \omega +f\right)  & 0 \\ 
0 & 0 & 1%
\end{array}%
\right) ,  \label{B_Mat}
\end{equation}%
% \begin{eqnarray}
% \mathbfss{D}\left( \Omega \right)  &=&\left( 
% \begin{array}{ccc}
% \cos \Omega  & -\sin \Omega  & 0 \\ 
% \sin \Omega  & \cos \Omega  & 0 \\ 
% 0 & 0 & 1%
% \end{array}%
% \right) ,  \label{D_Mat} \\
% \mathbfss{C}\left( i\right)  &=&\left( 
% \begin{array}{ccc}
% 1 & 0 & 0 \\ 
% 0 & \cos i & -\sin i \\ 
% 0 & \sin i & \cos i%
% \end{array}%
% \right) ,  \label{C_Mat} \\
% \mathbfss{B}\left( \omega +f\right)  &=&\left( 
% \begin{array}{ccc}
% \cos \left( \omega +f\right)  & -\sin \left( \omega +f\right)  & 0 \\ 
% \sin \left( \omega +f\right)  & \cos \left( \omega +f\right)  & 0 \\ 
% 0 & 0 & 1%
% \end{array}%
% \right) ,  \label{B_Mat}
% \end{eqnarray}%
% \begin{equation}
% \mathbfss{D}\left( \Omega \right)  =\left( 
% \begin{array}{ccc}
% \cos \Omega  & -\sin \Omega  & 0 \\ 
% \sin \Omega  & \cos \Omega  & 0 \\ 
% 0 & 0 & 1%
% \end{array}%
% \right) , \hspace{0.2cm}
% \mathbfss{C}\left( i\right) = \left( 
% \begin{array}{ccc}
% 1 & 0 & 0 \\ 
% 0 & \cos i & -\sin i \\ 
% 0 & \sin i & \cos i%
% \end{array}%
% \right) , \hspace{0.2cm}
% \mathbfss{B}\left( \omega +f\right)  =\left( 
% \begin{array}{ccc}
% \cos \left( \omega +f\right)  & -\sin \left( \omega +f\right)  & 0 \\ 
% \sin \left( \omega +f\right)  & \cos \left( \omega +f\right)  & 0 \\ 
% 0 & 0 & 1%
% \end{array}%
% \right) ,  \label{B_Mat}
% \end{equation}%
and where the node $\Omega $ and the argument of pericenter $\omega $ have
been updated using equations (\ref{TA_elorb}) or (\ref{MA_elorb}). 
\end{enumerate}

\subsection{Comparison of the reference orbits}\label{SPE_comparison}

The two reference trajectories defined above are clearly different. The main
difference between the two formulations is due to the equation of the
center, that is, to the periodic nature of the difference between the true
and the mean anomaly. Fundamentally this implies that the SPE\ in true
anomaly oscillates periodically with respect to the SPE in mean anomaly. 
\begin{comment}
The dependence on $f$ allows to the SPE\ in true anomaly to follow more closely
the real orbit as perturbed by $J_{2}.$ Numerical tests, in fact,
demonstrate that the difference between a numerically integrated orbit
perturbed by $J_{2}$ and the SPE in true anomaly oscillates with constant
amplitude. On the contrary the same comparison using the SPE in mean anomaly
shows a linear growth of the difference.
\end{comment}

\section{The SPE in the true anomaly }\label{Current_SPE}

This Section provides the formulation of the secularly precessing ellipse
with the true anomaly as the independent variable. We first review the
equations of motion developed by \cite{KynerBennett:1966} 
and then provide a more concise reformulation. 

\subsection{The Kyner and Bennett formulation}\label{KB_formulation}

The equations of motion can be derived easily by differentiating twice
equation (\ref{nominal_pos}) with respect to $t$ and taking into account
equation (\ref{TA_elorb}). We omit the derivation and give only the equations of
motion which describe the motion on SPE. The formulation is essentially the
same as the original formulation of Kyner and Bennett, that is%
% \begin{eqnarray}
% \bmath{\ddot{r}}+\dfrac{\mu _{0}\left( 1-\gamma \right) ^{2}}{r^{3}}\,%
% \bmath{r} &=&\mu _{0}\left( 1-\gamma \right) ^{2}\dfrac{\left( 1+e\cos
% f\right) }{r^{3}}  \label{KB_original_EOM} \\
% &&\times \left\{ \tau ^{2}\dfrac{d^{2}\mathbfss{D}}{d\Omega ^{2}}\mathbfss{CB}%
% +2\tau \left( 1+\eta \right) \dfrac{d\mathbfss{D}}{d\Omega }\mathbfss{C}\dfrac{d%
% \mathbfss{B}}{du}-\left( \eta ^{2}+2\eta \right) \mathbfss{R}\right\} 
% \bmath{r}_{rtn},  \nonumber
% \end{eqnarray}%
\begin{equation}
\bmath{\ddot{r}}+\dfrac{\mu _{0}\left( 1-\gamma \right) ^{2}}{r^{3}}\,%
\bmath{r} =\mu _{0}\left( 1-\gamma \right) ^{2}\dfrac{\left( 1+e\cos
f\right) }{r^{3}}  
\left\{ \tau ^{2}\dfrac{d^{2}\mathbfss{D}}{d\Omega ^{2}}\mathbfss{CB}%
+2\tau \left( 1+\eta \right) \dfrac{d\mathbfss{D}}{d\Omega }\mathbfss{C}\dfrac{d%
\mathbfss{B}}{du}-\left( \eta ^{2}+2\eta \right) \mathbfss{R}\right\} 
\bmath{r}_{rtn},  \label{KB_original_EOM}
\end{equation}%
the only differences being that we have factored $\left( 1-\gamma \right)
^{2}$ on the right hand side and summed the Newtonian term with the last
term on the right hand side of the original K\&B\ equation (12). Note that
here we have introduced the argument of latitude $u=\omega +f.$ The true
anomaly appears explicitly in the equations of motion (\ref{KB_original_EOM}%
)\ through the rotation matrices $\mathbfss{B}$ and $\mathbfss{D}$ (equations \ref{B_Mat} and 
\ref{D_Mat} respectively). It is thus
necessary to compute $f$ either from the analytic procedure outlined
previously, or by numerically integrating the auxiliary equation%
\begin{equation}
\dot{f}=\dfrac{n_{0}\left( 1-\gamma \right) }{\left( 1-e^{2}\right) ^{3/2}}%
\left( 1+e\cos f\right) ^{2}.  \label{KB_original_true_anom}
\end{equation}%
simultaneously with (\ref{KB_original_EOM}). Note that although the
perturbed mean motion $n$ appears, this equation is independent of the
radius vector $r.$ The term $1+e\cos f$ in equations (\ref{KB_original_EOM})
cannot be replaced with its Keplerian equivalent $a(1-e^{2})/r$ unless the
radius vector appearing at the denominator be computed by its Keplerian definition---which means going
back to using the true anomaly again, or using the result from the
integration of (\ref{KB_original_true_anom}). 

\subsection{A reformulation of the Kyner and Bennett approach}\label{KB_reformulation}

In this section we provide the derivation of an alternative form of the
equations of motion (\ref{KB_original_EOM}). 
Starting from the kinematic representation (\ref{nominal_pos}) of
the SPE we take the second derivative with respect to the time to obtain 
\begin{equation}
\frac{d^{2}\bmath{r}}{dt^{2}}=\left( \frac{df}{dt}\right) ^{2}\frac{%
d^{2}\mathbfss{R}}{df^{2}}\bmath{r}_{rtn}+\frac{d\mathbfss{R}}{df}\left[ 
\frac{d^{2}f}{dt^{2}}\bmath{r}_{rtn}+2\frac{df}{dt}\bmath{\dot{r}}%
_{rtn}\right] +\,\mathbfss{R}\,\bmath{\ddot{r}}_{rtn}.  \label{sec_der_KB}
\end{equation}%
In keeping with the Kyner \& Bennett approach (equations (\ref{TA_elorb})
and (\ref{TA_rates tau})-(\ref{TA gamma})), every term is now reshaped into
a form based on the true anomaly.

For the scalar components $\dot{r}$ and $\ddot{r}$ we need the first and second time derivatives of
both the radius vector and the true anomaly. From (\ref%
{KB_original_true_anom}) we immediately find%
\begin{equation}
\dot{f}=n\sqrt{ 1-e^{2} } \, \frac{a^{2}}{r^{2}},
\label{first_der_true_A}
\end{equation}%
with $n$ given by (\ref{n}), and subsequently%
\begin{equation}
\ddot{f}=-2\mu \frac{e\sin f}{r^{3}},  \label{sec_der_true_A}
\end{equation}%
where the perturbed gravitational parameter $\mu $ is given by%
\begin{equation}
\mu =n_{0}^{2}\left( 1-\gamma \right) ^{2}a^{3}.  \label{mu}
\end{equation}%
From the second of equations (\ref{pos_vel}) follows that 
\begin{equation}
\frac{dr}{dt}=\frac{na}{\sqrt{1-e^{2}}}e\sin f,  \label{first_der_r}
\end{equation}%
which can be differentiated to yield 
\begin{equation}
\frac{d^{2}r}{dt^{2}}=\mu \frac{e\cos f}{r^{2}}.  \label{sec_der_r}
\end{equation}%
For the matrix terms we need the derivatives of the rotation matrix $\mathbfss{%
R}$ with respect to true anomaly $f.$ These are easily computed from (\ref%
{R_Mat}) and with the help of (\ref{TA_elorb}) they can be written as 
\begin{eqnarray}
\frac{d\mathbfss{R}}{df} &=&\tau \frac{d\mathbfss{D}}{d\Omega }\mathbfss{CB}%
+\left( 1+\eta \right) \mathbfss{DC}\frac{d\mathbfss{B}}{du},
\label{first_der_R_Mat} \\
\frac{d^{2}\mathbfss{R}}{df^{2}} &=&\tau ^{2}\frac{d^{2}\mathbfss{D}}{d\Omega
^{2}}\mathbfss{CB}+2\tau \left( 1+\eta \right) \frac{d\mathbfss{D}}{d\Omega }%
\mathbfss{C}\frac{d\mathbfss{B}}{du}+\left( 1+\eta \right) ^{2}\mathbfss{DC}\frac{%
d^{2}\mathbfss{B}}{du^{2}}.  \label{sec_der_R_Mat}
\end{eqnarray}%
Direct use of these expressions clearly leads to the original K\&B
formulation. Our purpose here is instead to reformulate equations (\ref%
{first_der_R_Mat}) and (\ref{sec_der_R_Mat}) through the introduction of the
matrix operators $\mathcal{D}_{f}=d/df$ and $\mathcal{D}%
_{f}^{2}=d^{2}/df^{2}.$ It is shown in Appendix \ref{matrix_manipulation}
that%
\begin{eqnarray}
\mathcal{D}_{f} &=&\tau \mathbfss{V}+\left( 1+\eta \right) \mathbfss{H},
\label{D_f} \\
\mathcal{D}_{f}^{2} &=&\tau ^{2}\mathbfss{K}+2\tau \left( 1+\eta \right) 
\mathbfss{N}-\left( 1+\eta \right) ^{2}\mathbfss{I},  \label{D2_f}
\end{eqnarray}%
where $\mathbfss{V},$ $\mathbfss{H},$ $\mathbfss{K}$ and $\mathbfss{N}$ are matrices
defined in (\ref{V_Mat}), (\ref{H_Mat}), (\ref{K_mat}) and (\ref{N_mat}),
respectively, and where $\mathbfss{I}$ is the identity matrix of order three.
Then%
\begin{eqnarray}
\frac{d\mathbfss{R}}{df} &=&\mathcal{D}_{f}\mathbfss{R}=\left[ \tau \mathbfss{V}%
+\left( 1+\eta \right) \mathbfss{H}\right] \mathbfss{R}, \\
\frac{d^{2}\mathbfss{R}}{df^{2}} &=&\mathcal{D}_{f}^{2}\mathbfss{R}=\left[ \tau
^{2}\mathbfss{K}+2\tau \left( 1+\eta \right) \mathbfss{N}-\left( 1+\eta \right)
^{2}\right] \mathbfss{R},
\end{eqnarray}%
and 
\begin{eqnarray}
( \mathcal{D}_{f}\mathbfss{R}) \bmath{r}_{rtn} &=&\left[
\tau \mathbfss{V}+\left( 1+\eta \right) \mathbfss{H}\right] \bmath{r},
\label{first_der_rvet} \\
( \mathcal{D}_{f}^{2}\mathbfss{R} ) \bmath{r}_{rtn} &=&\left[
\tau ^{2}\mathbfss{K}+2\tau \left( 1+\eta \right) \mathbfss{N}-\left( 1+\eta
\right) ^{2}\right] \bmath{r}.  \label{second_der_rvet}
\end{eqnarray}

Substituting equations (\ref{first_der_true_A}), (\ref{sec_der_true_A}), (%
\ref{first_der_r}), (\ref{sec_der_r}), (\ref{first_der_rvet}) and (\ref%
{second_der_rvet}) into (\ref{sec_der_KB}) we find%
\begin{equation}
\bmath{\ddot{r}}=n^{2} ( 1-e^{2} ) \frac{a^{4}}{r^{4}}\left[
\tau ^{2}\mathbfss{K}+2\tau \left( 1+\eta \right) \mathbfss{N}-\left( 1+\eta
\right) ^{2}\right] \bmath{r}+\mu \frac{e\cos f}{r^{3}}\bmath{r}.
\end{equation}%
Notice that in performing the substitutions all traces of the first
derivative $D_{f}\mathbfss{R}$ have been lost. Adding and subtracting the term 
$\left( \mu /r^{3}\right) \bmath{r},$ with $\mu $ from (\ref{mu}), on
the right hand side yields 
\begin{equation}
\bmath{\ddot{r}}+\dfrac{\mu }{r^{3}}\bmath{r}=n^{2}\left(
1-e^{2}\right) \dfrac{a^{4}}{r^{4}}\left\{ \left[ 1-\left( 1+\eta \right)
^{2}\right] \mathbfss{I}+\tau ^{2}\mathbfss{K}+2\tau \left( 1+\eta \right) 
\mathbfss{N}\right\} \bmath{r}.
\end{equation}%
This equation can now be put in the compact form%
% \begin{equation}
% \fbox{$\bmath{\ddot{r}}+\dfrac{\mu }{r^{3}}\bmath{r}=\dfrac{\mu p}{%
% r^{4}}\mathbfss{S\,}\bmath{r},$}  \label{KB_Eom_reformulated}
% \end{equation}%
\begin{equation}
\fbox{$\bmath{\ddot{r}}+\displaystyle\dfrac{\mu }{r^{3}}\bmath{r}=\displaystyle\dfrac{\mu p}{%
r^{4}}\mathbfss{S\,}\bmath{r},$}  \label{KB_Eom_reformulated}
\end{equation}%
% % \begin{equation}
% % \fbox{
% %  \begin{minipage}{2.5cm}
% % \vspace{-0.2cm}
% %   \begin{eqnarray*}
% % \bmath{\ddot{r}}+\dfrac{\mu }{r^{3}}\bmath{r} = \dfrac{\mu p}{r^{4}}\mathbfss{S\,}\bmath{r},
% %   \end{eqnarray*}
% %  \end{minipage}
% % } \label{KB_Eom_reformulated}
% % \end{equation}
where $p=a\left( 1-e^{2}\right) $ is the orbital semi-latus rectum, or
parameter of the ellipse, and the matrix $\mathbfss{S}$ has been defined as%
\begin{equation}
\mathbfss{S}=\left( 
\begin{array}{ccc}
A & 0 & B\sin \Omega  \\ 
0 & A & B\cos \Omega  \\ 
0 & 0 & C%
\end{array}%
\right) \,,
\end{equation}%
with the constants 
\begin{eqnarray}
A &=&1-\left( 1+\eta \right) ^{2}-\tau \left\{ \tau +2\left( 1+\eta \right)
\cos i\right\} , \\
B &=&2\tau \left( 1+\eta \right) \sin i, \\
C &=&1-\left( 1+\eta \right) ^{2}.
\end{eqnarray}

In component form, with $\bmath{r}=\left( x,y,z\right) ^{T},$ the
reformulated equations of motion of Kyner \&\ Bennett read%
\begin{eqnarray}
\ddot{x}+\dfrac{\mu }{r^{3}}x &=&\dfrac{\mu p}{r^{4}}\left\{ \left[ 1-\left(
1+\eta \right) ^{2}-\tau \left\{ \tau +2\left( 1+\eta \right) \cos i\right\} %
\right] x+\left[ 2\tau \left( 1+\eta \right) \sin i\sin \Omega \right]
z\right\} ,  \nonumber \\
\ddot{y}+\dfrac{\mu }{r^{3}}y &=&\dfrac{\mu p}{r^{4}}\left\{ \left[ 1-\left(
1+\eta \right) ^{2}-\tau \left\{ \tau +2\left( 1+\eta \right) \cos i\right\} %
\right] y+\left[ 2\tau \left( 1+\eta \right) \sin i\cos \Omega \right]
z\right\} ,  \nonumber \\
\ddot{z}+\dfrac{\mu }{r^{3}}z &=&\dfrac{\mu p}{r^{4}}\left[ 1-\left( 1+\eta
\right) ^{2}\right] z.
\end{eqnarray}

Equations (\ref{KB_Eom_reformulated}), although highly simplified with respect to 
equations (\ref{KB_original_EOM})---note the absence of matrix 
multiplications---still need to be supported by equation (\ref%
{KB_original_true_anom}), since true anomaly appears implicitly through the
node $\Omega $ (see equation (\ref{TA_elorb})). 

\section{The SPE in the mean anomaly }\label{New_SPE}

This Section provides the formulation of the secularly precessing ellipse
with the time, or the mean anomaly, as the independent variable. We
review the model due to \cite{Escobal1965} and develop alternative
formulations following essentially the same original line of development,
but which are more concise for implementation.

\subsection{The Escobal formulation}\label{Escobal_formulation}

The procedure to obtain the equations of motion is well described in \cite%
{Escobal1965}. Here we give only the final result and the necessary
auxiliary equation in case it is desired to fully integrate the equations of
motion as a quick alternative to following the analytical update of the
eccentric anomaly as envisioned in \cite{Escobal1965}. The equations of
motion are derived from the decomposition of the radius vector 
\begin{equation}
\bmath{r}=X\bmath{P}+Y\bmath{Q,}
\end{equation}%
in the Laplace reference frame, where $\bmath{P}$ is the
Hermann-Jacobi-Laplace unit vector and $\bmath{Q}$ is a unit vector
normal to $\bmath{P}$ in the direction of increasing anomaly. The unit
vectors $\bmath{P}$ and $\bmath{Q}$ are themselves defined as 
\begin{equation}
\hspace{-0.2cm}
\begin{array}{ll}
P_{x}=\cos \omega \cos \Omega -\sin \omega \sin \Omega \cos i,\qquad & 
Q_{x}=-\sin \omega \cos \Omega -\cos \omega \sin \Omega \cos i, \\ 
P_{y}=\cos \omega \sin \Omega +\sin \omega \cos \Omega \cos i,\qquad & 
Q_{y}=-\sin \omega \sin \Omega +\cos \omega \cos \Omega \cos i, \\ 
P_{z}=\sin \omega \sin i, & Q_{z}=\cos \omega \sin i.%
\end{array}%
\,  \label{PQ}
\end{equation}

The acceleration is then 
\begin{equation}
\bmath{\ddot{r}}+\frac{\bar{\mu}}{r^{3}}\bmath{r}=2\dot{X}%
\bmath{\dot{P}}+2\dot{Y}\bmath{\dot{Q}}+X\bmath{\ddot{P}}+Y%
\bmath{\ddot{Q}}\,,  \label{Escobal_EOM}
\end{equation}%
where the modified gravitational parameter $\bar{\mu}$ is given by%
\begin{equation}
\bar{\mu}=\bar{n}^{2}a^{3},  \label{mu_bar}
\end{equation}%
and the first and second derivatives of $\bmath{P}=\left(
P_{x},P_{y},P_{z}\right) ^{T}$ and $\bmath{Q}=\left(
Q_{x},Q_{y},Q_{z}\right) ^{T}$ are given by%
\begin{equation}
\hspace{-0.2cm}
\begin{array}{ll}
\dot{P}_{x}=-\dot{\Omega}P_{y}+\dot{\omega}Q_{x},\qquad  & \dot{Q}_{x}=-\dot{%
\Omega}Q_{y}-\dot{\omega}P_{x}, \\ 
\dot{P}_{y}=\dot{\Omega}P_{x}+\dot{\omega}Q_{y}, & \dot{Q}_{y}=\dot{\Omega}%
Q_{x}-\dot{\omega}P_{y}, \\ 
\dot{P}_{z}=\dot{\omega}Q_{z}, & \dot{Q}_{z}=-\dot{\omega}P_{z},%
\end{array}%
\,  \label{PQ_first_der}
\end{equation}%
and%
\begin{equation}
\hspace{-0.2cm}
\begin{array}{ll}
\ddot{P}_{x}=-\left( \dot{\Omega}^{2}+\dot{\omega}^{2}\right) P_{x}-2\dot{%
\Omega}\dot{\omega}Q_{y},\qquad  & \ddot{Q}_{x}=-\left( \dot{\Omega}^{2}+%
\dot{\omega}^{2}\right) Q_{x}+2\dot{\Omega}\dot{\omega}P_{y}, \\ 
\ddot{P}_{y}=-\left( \dot{\Omega}^{2}+\dot{\omega}^{2}\right) P_{y}+2\dot{%
\Omega}\dot{\omega}Q_{x}, & \ddot{Q}_{y}=-\left( \dot{\Omega}^{2}+\dot{\omega%
}^{2}\right) Q_{y}-2\dot{\Omega}\dot{\omega}P_{x}, \\ 
\ddot{P}_{z}=-\dot{\omega}^{2}P_{z}, & \ddot{Q}_{z}=-\dot{\omega}^{2}Q_{z}.%
\end{array}%
\,  \label{PQ_sec_der}
\end{equation}%
Finally, $X,$ $\dot{X},$ $Y$ and $\dot{Y}$ are defined either in terms of
the eccentric anomaly $E$ or in terms of the true anomaly $f$ as 
\begin{equation}
\hspace{-0.2cm}
\begin{array}{ll}
X=a\left( \cos E-e\right) =r\cos f, & \dot{X}=-\displaystyle\dfrac{\bar{n}a^{2}}{r}\sin
E=-\displaystyle\dfrac{\bar{n}a}{\sqrt{1-e^{2}}}\sin f, \\ 
Y=a\sqrt{1-e^{2}}\sin E=r\sin f,\qquad  & \dot{Y}=\displaystyle\dfrac{\bar{n}a^{2}}{r}%
\sqrt{1-e^{2}}\cos E=\displaystyle\dfrac{\bar{n}a\left( e+\cos f\right) }{\sqrt{1-e^{2}}},%
\end{array}
\label{XXdotYYdot}
\end{equation}%
where the perturbed mean motion $\bar{n}$ is given by equation (\ref{n_bar}%
). As in the true anomaly-based Kyner and Bennett formulation the number of
first order differential equation to integrate is effectively seven, that
is, six equations for the state vector and one equation (uncoupled with the
state) for eccentric anomaly%
\begin{equation}
\frac{dE}{dt}=\frac{\bar{n}}{1-e\cos E},  \label{Escobal_eom_ecc_anom}
\end{equation}%
or the true anomaly%
\begin{equation}
\frac{df}{dt}=\dfrac{\bar{n}}{\left( 1-e^{2}\right) ^{3/2}}\left( 1+e\cos
f\right) ^{2}.  \label{Escobal_eom_true_anom}
\end{equation}%
Note that the equation for the anomaly is not redundant, but provides
necessary information for the integration of the differential equations (\ref%
{Escobal_EOM}).

\subsection{A reformulation of the Escobal approach}\label{Escobal_reformulation}

A simpler form of Escobal's equations of motion can be obtained by
resolving Eq. (\ref{Escobal_EOM}) in inertial axes and substituting in it
Eqs. (\ref{PQ_first_der}), (\ref{PQ_sec_der}), (\ref{PQ}) and (\ref%
{XXdotYYdot}) for the true anomaly to obtain
\begin{equation}
\bmath{\ddot{r}}+\frac{\bar{\mu}}{r^{3}}\bmath{r}=-e\bmath{A}%
-\left\{ \bmath{B}r+\bmath{A}\right\} \cos f-\left\{ \bmath{C}%
r+\bmath{D}\right\} \sin f,  \label{Escobal_EOM_reformulated_vectors}
\end{equation}%
where the components of the vectors $\bmath{A},$ $\bmath{B},$ $%
\bmath{C}$ and $\bmath{D}$ are%
\begin{equation}
\hspace{-0.2cm}
\begin{array}{ll}
A_{x}=\displaystyle\frac{2\bar{n}a}{\sqrt{1-e^{2}}}\left( \dot{\omega}P_{x}+\dot{\Omega}%
Q_{y}\right) ,\qquad  & B_{x}=\left( \dot{\Omega}^{2}+\dot{\omega}%
^{2}\right) P_{x}+2\dot{\Omega}\dot{\omega}Q_{y}, \\ 
A_{y}=\displaystyle\frac{2\bar{n}a}{\sqrt{1-e^{2}}}\left( \dot{\omega}P_{y}-\dot{\Omega}%
Q_{x}\right) ,\qquad  & B_{y}=\left( \dot{\Omega}^{2}+\dot{\omega}%
^{2}\right) P_{y}-2\dot{\Omega}\dot{\omega}Q_{x}, \\ 
A_{z}=\displaystyle\frac{2\bar{n}a}{\sqrt{1-e^{2}}}\dot{\omega}P_{z}, & B_{z}=\dot{\omega}%
^{2}P_{z},\qquad  \\ 
C_{x}=-2\dot{\Omega}\dot{\omega}P_{y}+\left( \dot{\Omega}^{2}+\dot{\omega}%
^{2}\right) Q_{x},\qquad  & D_{x}=\displaystyle\frac{2\bar{n}a}{\sqrt{1-e^{2}}}\left[ -%
\dot{\Omega}P_{y}+\dot{\omega}Q_{x}\right] , \\ 
C_{y}=2\dot{\Omega}\dot{\omega}P_{x}+\left( \dot{\Omega}^{2}+\dot{\omega}%
^{2}\right) Q_{y},\qquad  & D_{y}=\displaystyle\frac{2\bar{n}a}{\sqrt{1-e^{2}}}\left[ 
\dot{\Omega}P_{x}+\dot{\omega}Q_{y}\right] , \\ 
C_{z}=\dot{\omega}^{2}Q_{z}, & D_{z}=\displaystyle\frac{2\bar{n}a}{\sqrt{1-e^{2}}}\dot{%
\omega}Q_{z}.%
\end{array}
\label{ABCD}
\end{equation}

Equations (\ref%
{Escobal_EOM_reformulated_vectors}) are written explicitly in the true
anomaly, and this requires that for their numerical integration they be
supplemented with equation (\ref{Escobal_eom_true_anom}).

Alternatively, one can choose to work with the eccentric anomaly, in which
case the equations of motion read
\begin{equation}
\bmath{\ddot{r}}+\frac{\bar{\mu}}{r^{3}}\bmath{r}=e\bmath{B}%
^{\prime }-\left\{ \bmath{A}^{\prime }\frac{1}{r}+\bmath{B}%
^{\prime }\right\} \cos E-\left\{ \bmath{C}^{\prime }+\bmath{D}%
^{\prime }\frac{1}{r}\right\} \sin E,
\end{equation}%
and the components of the several vector coefficient are
\begin{equation}
\hspace{-0.2cm}
\begin{array}{ll}
\bmath{A}^{\prime }=a\left( 1-e^{2}\right) \bmath{A},\hspace{0.9cm}
& \bmath{B}^{\prime }=a\bmath{B}, \\ 
\bmath{C}^{\prime }=a\sqrt{1-e^{2}}\bmath{C}, & \bmath{D}%
^{\prime }=a\sqrt{1-e^{2}}\bmath{D}.%
\end{array}
\label{A'B'C'D'2}
\end{equation}

\section{A new formulation of the SPE in the mean anomaly $M$}\label{New_SPE_mean}

In this section we give the detailed procedure we have followed to obtain a
new form of the equations of motion respresenting the SPE in the mean
anomaly. Starting from the kinematic representation of SPE (\ref{nominal_pos}%
) and taking the second derivative with respect to the time we have%
\begin{equation}
\frac{d^{2}\bmath{r}}{dt^{2}}=\mathbfss{R}\frac{d^{2}\bmath{r}_{rtn}%
}{dt^{2}}+2\frac{d\mathbfss{R}}{dt}\frac{d\bmath{r}_{rtn}}{dt}+\frac{d^{2}%
\mathbfss{R}}{dt^{2}}\bmath{r}_{rtn}\,,  \label{sec_der_MA}
\end{equation}%
which are to be used in connection with equations (\ref{MA_rates N}), (\ref{MA_rates omega}%
), (\ref{MA gamma}) and (\ref{MA_elorb}) that give the secular rates, the constant $%
\bar{\gamma}$ and the evolution of the orbital elements with respect to the mean
anomaly.

The time derivatives of the rotation matrix $\mathbfss{R}$ can be easily
computed from (\ref{R_Mat}) recalling (\ref{MA_elorb}). As in the case of
the K\&B reformulation of Section \ref{KB_reformulation}, the strategy is to
define appropriate operators $\mathcal{D}_{t}=d/dt$ and $\mathcal{D}%
_{t}^{2}=d^{2}/dt^{2}$ acting on the rotation matrix $\mathbfss{R.}$ In
Appendix \ref{matrix_manipulation_MA} it is shown that these operators can
be expressed as%
\begin{eqnarray}
\mathcal{D}_{t} &=&(\dot{\omega}+\dot{f})\mathbfss{H}+\dot{\Omega}\mathbfss{V},
\label{D_t} \\
\mathcal{D}_{t}^{2} &=&(\dot{\omega}+\dot{f})^{2}\mathbfss{H}^{2}+2\dot{\Omega}%
(\dot{\omega}+\dot{f})\mathbfss{N}+\dot{\Omega}^{2}\mathbfss{K}+\ddot{f}\mathbfss{H%
}.  \label{D2_t}
\end{eqnarray}

If we now recall that the use of the time as independent variable implies
from the second of (\ref{pos_vel}) that%
\begin{eqnarray}
\dot{r} &=&\frac{\bar{n}a}{\sqrt{1-e^{2}}}e\sin f,  \label{r_dot MA} \\
\ddot{r} &=&\bar{\mu}\frac{e\cos f}{r^{2}},  \label{r_dotdot MA}
\end{eqnarray}%
with $\bar{\mu}$ given by (\ref{mu_bar}), we can write%
\begin{eqnarray}
\frac{d\bmath{r}_{rtn}}{dt} &=&\frac{\bar{n}ae\sin f}{r\sqrt{1-e^{2}}}%
\bmath{r}_{rtn},  \label{d_r_rtn/dt} \\
\frac{d^{2}\bmath{r}_{rtn}}{dt^{2}} &=&\bar{\mu}\frac{e\cos f}{r^{3}}%
\bmath{r}_{rtn},  \label{d2_r_rtn/dt}
\end{eqnarray}%
and substitute (\ref{D_t}), (\ref{D2_t}), (\ref{d_r_rtn/dt}) and (\ref%
{d2_r_rtn/dt}) into (\ref{sec_der_MA}), we get the equations of motion in
the form\qquad\ 
% \begin{eqnarray}
% \bmath{\ddot{r}} &=&\bar{\mu}\frac{e\cos f}{r^{3}}\bmath{r}+2\frac{%
% \bar{n}ae\sin f}{r\sqrt{1-e^{2}}}\left[ (\dot{\omega}+\dot{f})\mathbfss{H}+%
% \dot{\Omega}\mathbfss{V}\right] \bmath{r}  \nonumber \\
% &&+\left[ (\dot{\omega}+\dot{f})^{2}\mathbfss{H}^{2}+2\dot{\Omega}(\dot{\omega}%
% +\dot{f})\mathbfss{N}+\dot{\Omega}^{2}\mathbfss{K}+\ddot{f}\mathbfss{H}\right] 
% \bmath{r}.  \label{d2r_dt2_1}
% \end{eqnarray}%
\begin{equation}
\bmath{\ddot{r}} =\bar{\mu}\frac{e\cos f}{r^{3}}\bmath{r}+2\frac{%
\bar{n}ae\sin f}{r\sqrt{1-e^{2}}}\left[ (\dot{\omega}+\dot{f})\mathbfss{H}+%
\dot{\Omega}\mathbfss{V}\right] \bmath{r}  
+\left[ (\dot{\omega}+\dot{f})^{2}\mathbfss{H}^{2}+2\dot{\Omega}(\dot{\omega}%
+\dot{f})\mathbfss{N}+\dot{\Omega}^{2}\mathbfss{K}+\ddot{f}\mathbfss{H}\right] 
\bmath{r}.  \label{d2r_dt2_1}
\end{equation}%
We now need expressions for the first and the second time derivatives of the
true anomaly. These can be obtained from (\ref{Escobal_eom_true_anom}) and
put in the form%
\begin{eqnarray}
\dot{f} &=&\bar{n}\sqrt{ 1-e^{2} } \, \frac{a^{2}}{r^{2}},
\label{f_dot MA} \\
\ddot{f} &=&2\frac{\bar{n}ae}{r\sqrt{1-e^{2}}}\dot{f}\sin f.
\label{f_dotdot MA}
\end{eqnarray}%
We thus easily see that in (\ref{d2r_dt2_1}) the term with $\dot{f}\mathbfss{H}
$ as a factor cancels with the term $\ddot{f}\mathbfss{H.}$ If at the same
time we add the modified Keplerian term $\bar{\mu}\bmath{r}/r^{3}$ to
both the left and right hand sides, the equations of motion become%
% \begin{eqnarray}
% \bmath{\ddot{r}}+\frac{\bar{\mu}}{r^{3}}\bmath{r} &=&\bar{\mu}%
% \frac{1+e\cos f}{r^{3}}\bmath{r}+2\frac{\bar{n}ae\sin f}{r\sqrt{1-e^{2}}%
% }\left( \dot{\omega}\mathbfss{H}+\dot{\Omega}\mathbfss{V}\right) \bmath{r} 
% \nonumber \\
% &&+\left[ (\dot{\omega}+\dot{f})^{2}\mathbfss{H}^{2}+2\dot{\Omega}(\dot{\omega}%
% +\dot{f})\mathbfss{N}+\dot{\Omega}^{2}\mathbfss{K}\right] \bmath{r}.
% \label{EoM_new1}
% \end{eqnarray}%
\begin{equation}
\bmath{\ddot{r}}+\frac{\bar{\mu}}{r^{3}}\bmath{r} =\bar{\mu}%
\frac{1+e\cos f}{r^{3}}\bmath{r}+2\frac{\bar{n}ae\sin f}{r\sqrt{1-e^{2}}%
}\left( \dot{\omega}\mathbfss{H}+\dot{\Omega}\mathbfss{V}\right) \bmath{r} 
+\left[ (\dot{\omega}+\dot{f})^{2}\mathbfss{H}^{2}+2\dot{\Omega}(\dot{\omega}%
+\dot{f})\mathbfss{N}+\dot{\Omega}^{2}\mathbfss{K}\right] \bmath{r}.
\label{EoM_new1}
\end{equation}%

Considering that, on the basis of the relationship (\ref{H2R}), the action
of $\mathbfss{H}^{2}$ on $\bmath{r}$ is simply to reverse its sign, 
\begin{equation}
\mathbfss{H}^{2}\bmath{r}=\mathbfss{H}^{2}\mathbfss{R}\bmath{r}_{rtn}=%
\mathbfss{RK}\bmath{r}_{rtn}=-\mathbfss{R}\bmath{r}_{rtn}=-\bmath{%
r},
\end{equation}%
we can write 
\begin{equation}
(\dot{f})^{2}\mathbfss{H}^{2}\bmath{r}=-\bar{\mu}\frac{p}{r^{4}}%
\bmath{r}=-\bar{\mu}\frac{1+e\cos f}{r^{3}}\bmath{r},
\end{equation}%
and thus eliminate the first term on the right hand side of (\ref{EoM_new1})
with the term proportional to $(\dot{f})^{2}.$ We thus obtain 
% \begin{eqnarray}
% \bmath{\ddot{r}}+\frac{\bar{\mu}}{r^{3}}\bmath{r} &=&2\frac{\bar{n}%
% ae\sin f}{r\sqrt{1-e^{2}}}\left( \dot{\omega}\mathbfss{H}+\dot{\Omega}\mathbfss{V%
% }\right) \bmath{r}  \label{EoM_new2} \\
% &&+\left[ 2\dot{\Omega}(\dot{\omega}+\dot{f})\mathbfss{N}+\dot{\Omega}^{2}%
% \mathbfss{K}\right] \bmath{r}-(\dot{\omega}^{2}+2\dot{\omega}\dot{f})%
% \bmath{r}.  \nonumber
% \end{eqnarray}%
\begin{equation}
\bmath{\ddot{r}}+\frac{\bar{\mu}}{r^{3}}\bmath{r} =2\frac{\bar{n}%
ae\sin f}{r\sqrt{1-e^{2}}}\left( \dot{\omega}\mathbfss{H}+\dot{\Omega}\mathbfss{V%
}\right) \bmath{r}  +\left[ 2\dot{\Omega}(\dot{\omega}+\dot{f})\mathbfss{N}+\dot{\Omega}^{2}%
\mathbfss{K}\right] \bmath{r}-(\dot{\omega}^{2}+2\dot{\omega}\dot{f})%
\bmath{r}.  \label{EoM_new2} 
\end{equation}%
Collecting the terms depending on $\dot{f}$ and substituting $\dot{f}$ from (%
\ref{f_dot MA}) and eliminating $\bar{n}$ in favor of the modified area
integral 
\begin{equation}
\bar{h}=\bar{n}a^{2}\sqrt{1-e^{2}},  \label{h_bar}
\end{equation}%
we obtain 
\begin{equation}
\bmath{\ddot{r}}+\frac{\bar{\mu}}{r^{3}}\bmath{r}=\left( \dot{%
\Omega}^{2}\mathbfss{K}+2\dot{\Omega}\dot{\omega}\mathbfss{N}-\dot{\omega}%
^{2}\mathbfss{I}\right) \bmath{r}+2\frac{\bar{h}e}{rp}\sin f\left( \dot{\omega}%
\mathbfss{H}+\dot{\Omega}\mathbfss{V}\right) \bmath{r}+2\frac{\bar{h}}{r^{2}%
}\left( \dot{\Omega}\mathbfss{N}-\dot{\omega}\mathbfss{I}\right) \bmath{r}.
\label{EoM_new4}
\end{equation}%
If we now define%
\begin{eqnarray}
\mathbfss{E} &=&\dot{\Omega}^{2}\mathbfss{K}+2\dot{\Omega}\dot{\omega}\mathbfss{N}-%
\dot{\omega}^{2},  \label{E_mtx} \\
\mathbfss{F} &=&\dot{\omega}\mathbfss{H}+\dot{\Omega}\mathbfss{V},  \label{F_mtx}
\\
\mathbfss{G} &=&\dot{\Omega}\mathbfss{N}-\dot{\omega}\mathbfss{I},  \label{G_mtx}
\end{eqnarray}%
the equations of motion assume the more compact form%
\begin{equation}
\bmath{\ddot{r}}+\dfrac{\bar{\mu}}{r^{3}}\bmath{r}=\left( \mathbfss{E%
}+2\frac{\bar{h}e}{pr}\sin f\,\mathbfss{F}+2\dfrac{\bar{h}}{r^{2}}\mathbfss{G}%
\right) \bmath{r}.  \label{EoM_new5}
\end{equation}%
The matrices $\mathbfss{E}$ and $\mathbfss{G}$ are upper triangular, while
matrix $\mathbfss{F}$ is antisymetric. These three matrices are time-dependent
through the longitude of the node, and are given explicitly as%
\begin{eqnarray}
\mathbfss{E}\left( t\right)  &=&\left( 
\begin{array}{ccc}
-\left( \dot{\Omega}^{2}+2\dot{\Omega}\dot{\omega}\cos i+\dot{\omega}%
^{2}\right)  & 0 & 2\dot{\Omega}\dot{\omega}\sin i\sin \Omega \left(
t\right)  \\ 
0 & -\left( \dot{\Omega}^{2}+2\dot{\Omega}\dot{\omega}\cos i+\dot{\omega}%
^{2}\right)  & -2\dot{\Omega}\dot{\omega}\sin i\cos \Omega \left( t\right) 
\\ 
0 & 0 & -\dot{\omega}^{2}%
\end{array}%
\right) ,  \label{E_Mat} \\
\mathbfss{F}\left( t\right)  &=&\left( 
\begin{array}{ccc}
0 & -\left( \dot{\Omega}+\dot{\omega}\cos i\right)  & -\dot{\omega}\sin
i\cos \Omega \left( t\right)  \\ 
\left( \dot{\Omega}+\dot{\omega}\cos i\right)  & 0 & -\dot{\omega}\sin i\sin
\Omega \left( t\right)  \\ 
\dot{\omega}\sin i\cos \Omega \left( t\right)  & \dot{\omega}\sin i\sin
\Omega \left( t\right)  & 0%
\end{array}%
\right) =-\mathbfss{F}^{T}\,\left( t\right) ,  \label{F_Mat} \\
\mathbfss{G}\left( t\right)  &=&\left( 
\begin{array}{ccc}
-\left( \dot{\Omega}\cos i+\dot{\omega}\right)  & 0 & \dot{\Omega}\sin i\sin
\Omega \left( t\right)  \\ 
0 & -\left( \dot{\Omega}\cos i+\dot{\omega}\right)  & -\dot{\Omega}\sin
i\cos \Omega \left( t\right)  \\ 
0 & 0 & -\dot{\omega}%
\end{array}%
\right) .  \label{G_Mat}
\end{eqnarray}

It appears odd that the longitude of the pericenter does not appear
explicitly in the formulation. However, we need to recall that the explicit
dependence on time is restricted to the position of the orbital plane, which
depends on the inclination $i$ and the node $\Omega ,$ only the latter of
which is variable in our present model.

The final and crucial step is to recognize from (\ref{d_r_rtn/dt}) that 
\begin{equation}
\frac{dr}{dt}=\frac{\bar{h}e}{p}\sin f\,=\bmath{\dot{r}\cdot \hat{r}},
\end{equation}%
so that Eq. (\ref{EoM_new5}) can be reduced to the form%
% \begin{equation}
% \fbox{$\bmath{\ddot{r}}+\dfrac{\bar{\mu}}{r^{3}}\bmath{r}=\left[ 
% \mathbfss{E}+\dfrac{2}{r}\left( \bmath{\dot{r}\cdot \hat{r}}\right) 
% \mathbfss{F}+\dfrac{2\bar{h}}{r^{2}}\mathbfss{G}\right] \bmath{r}.$}
% \label{Eom_new_FINAL}
% \end{equation}%
\begin{equation}
\fbox{$\bmath{\ddot{r}}+\displaystyle\dfrac{\bar{\mu}}{r^{3}}\bmath{r}=\left[ 
\mathbfss{E}+\displaystyle\dfrac{2}{r}\left( \bmath{\dot{r}\cdot \hat{r}}\right) 
\mathbfss{F}+\displaystyle\dfrac{2\bar{h}}{r^{2}}\mathbfss{G}\right] \bmath{r}.$}
\label{Eom_new_FINAL}
\end{equation}%
The right-hand side of this equation is the explicit form of the disturbing acceleration
$\bmath{s}\left( \bmath{r},\bmath{\dot{r}},t\right)$ appearing in equation (\ref{Encke EoM 1}).
% % \begin{equation}
% % \fbox{
% %  \begin{minipage}{5.5cm}
% % \vspace{-0.2cm}
% %   \begin{eqnarray*}
% % \bmath{\ddot{r}}+\dfrac{\bar{\mu}}{r^{3}}\bmath{r}=\left[ 
% % \mathbfss{E}+\dfrac{2}{r}\left( \bmath{\dot{r}\cdot \hat{r}}\right) 
% % \mathbfss{F}+\dfrac{2\bar{h}}{r^{2}}\mathbfss{G}\right] \bmath{r}.
% %   \end{eqnarray*}
% %  \end{minipage}
% % } \label{Eom_new_FINAL}
% % \end{equation}
In component form, with $\bmath{r}=\left( x,y,z\right) ^{T}$, equation (\ref{Eom_new_FINAL}) reads
% \begin{eqnarray}
% \ddot{x}+\frac{\bar{\mu}}{r^{3}}x &=&-\left( \dot{\Omega}^{2}+2\dot{\Omega}%
% \dot{\omega}\cos i+\dot{\omega}^{2}\right) x+\left( 2\dot{\Omega}\dot{\omega}%
% \sin i\sin \Omega \right) z  \nonumber \\
% &&+2\frac{x\dot{x}+y\dot{y}+z\dot{z}}{r^{2}}\left[ -\left( \dot{\Omega}+\dot{%
% \omega}\cos i\right) y-\left( \dot{\omega}\sin i\cos \Omega \right) z\right] 
% \nonumber \\
% &&+2\frac{\bar{h}}{r^{2}}\left[ -\left( \dot{\Omega}\cos i+\dot{\omega}%
% \right) x+\left( \dot{\Omega}\sin i\sin \Omega \right) z\right] ,
% \label{EoM_NEW_x} \\
% \ddot{y}+\frac{\bar{\mu}}{r^{3}}y &=&-\left( \dot{\Omega}^{2}+2\dot{\Omega}%
% \dot{\omega}\cos i+\dot{\omega}^{2}\right) y-\left( 2\dot{\Omega}\dot{\omega}%
% \sin i\cos \Omega \right) z  \nonumber \\
% &&+2\frac{x\dot{x}+y\dot{y}+z\dot{z}}{r^{2}}\left[ \left( \dot{\Omega}+\dot{%
% \omega}\cos i\right) x-\left( \dot{\omega}\sin i\sin \Omega \right) z\right] 
% \nonumber \\
% &&+2\frac{\bar{h}}{r^{2}}\left[ -\left( \dot{\Omega}\cos i+\dot{\omega}%
% \right) y-\left( \dot{\Omega}\sin i\cos \Omega \right) z\right] ,
% \label{EoM_NEW_y} \\
% \ddot{z}+\frac{\bar{\mu}}{r^{3}}z &=&2\frac{x\dot{x}+y\dot{y}+z\dot{z}}{r^{2}%
% }\left[ \left( \dot{\omega}\sin i\cos \Omega \right) x+\left( \dot{\omega}%
% \sin i\sin \Omega \right) y\right] -\dot{\omega}\left( \dot{\omega}+2\frac{%
% \bar{h}}{r^{2}}\right) z.  \label{EoM_NEW_z}
% \end{eqnarray}
\begin{eqnarray}
\ddot{x}+\frac{\bar{\mu}}{r^{3}}x &=&-\left( \dot{\Omega}^{2}+2\dot{\Omega}%
\dot{\omega}\cos i+\dot{\omega}^{2}\right) x+\left( 2\dot{\Omega}\dot{\omega}%
\sin i\sin \Omega \right) z  
+2\frac{x\dot{x}+y\dot{y}+z\dot{z}}{r^{2}}\left[ -\left( \dot{\Omega}+\dot{%
\omega}\cos i\right) y-\left( \dot{\omega}\sin i\cos \Omega \right) z\right] 
\nonumber \\
&&+2\frac{\bar{h}}{r^{2}}\left[ -\left( \dot{\Omega}\cos i+\dot{\omega}%
\right) x+\left( \dot{\Omega}\sin i\sin \Omega \right) z\right] ,
\label{EoM_NEW_x} \\
\ddot{y}+\frac{\bar{\mu}}{r^{3}}y &=&-\left( \dot{\Omega}^{2}+2\dot{\Omega}%
\dot{\omega}\cos i+\dot{\omega}^{2}\right) y-\left( 2\dot{\Omega}\dot{\omega}%
\sin i\cos \Omega \right) z 
+2\frac{x\dot{x}+y\dot{y}+z\dot{z}}{r^{2}}\left[ \left( \dot{\Omega}+\dot{%
\omega}\cos i\right) x-\left( \dot{\omega}\sin i\sin \Omega \right) z\right] 
\nonumber \\
&&+2\frac{\bar{h}}{r^{2}}\left[ -\left( \dot{\Omega}\cos i+\dot{\omega}%
\right) y-\left( \dot{\Omega}\sin i\cos \Omega \right) z\right] ,
\label{EoM_NEW_y} \\
\ddot{z}+\frac{\bar{\mu}}{r^{3}}z &=&2\frac{x\dot{x}+y\dot{y}+z\dot{z}}{r^{2}%
}\left[ \left( \dot{\omega}\sin i\cos \Omega \right) x+\left( \dot{\omega}%
\sin i\sin \Omega \right) y\right] -\dot{\omega}\left( \dot{\omega}+2\frac{%
\bar{h}}{r^{2}}\right) z.  \label{EoM_NEW_z}
\end{eqnarray}
This form of the equations of motion of the secularly precessing ellipse has
the benefit of being a \emph{minimal} and \emph{complete} set of equations. In fact Eq. (\ref%
{Eom_new_FINAL}) has coefficients which are all explicit functions of the
time and there is no longer any need for the supporting differential
equation (\ref{Escobal_eom_true_anom}) for the anomaly $f$.

\section{Discussion and Conclusions}\label{Conclusions}

The Kyner and Bennett and the Escobal equations of motion defining the
intermediary orbit as an ellipse subject to secular motion of its angular
elements have been reviewed. 
In both cases a more compact formulation has been developed and presented,
which is better suited for implementation. 
Escobal's equations have also been reformulated and extended to allow the 
use of either the true or the eccentric anomaly, in addition to the mean anomaly. 
In fact, both these classical formulations are {\lq}redundant,\rq 
in that they include one more differential equation than the minimum of six 
associated with the three degrees of freedom of the problem. 
The reason for this extra equation is due to the particular choice of variables,
which makes it necessary to propagate the perturbed anomaly, be it the eccentric
or the true anomaly, along with the dynamical variables.

The main contribution of the present work is the development of a novel
formulation of the equations of motion of the secularly precessing ellipse
that uses the time as the independent variable and requires no additional
equation to account for the evolution of the anomaly. 
The final equation has the very compact form of a Two-Body equation of motion perturbed 
by a time-dependent acceleration containing three terms of degree 0 and $\pm1$ in the radius vector.
The explicit dependence on the time is only due to the presence of the longitude of the node in the forcing terms.
It should be noted that the supporting equation for the anomaly cannot be simplified away 
for true anomaly-based theories like Kyner and Bennett's. 

The time-wise approach developed here can be used when it is desired to numerically verify
the analytical propagation of perturbations based on Kaula-type linear theories,
where the nominal trajectory is a secularly precessing orbit. 
In particular, it can be used to verify the perturbation spectrum.
It can also be used when analyzing first-order perturbation effects on orbital arcs, 
or ephemerides, estimated from observational data. In that case the secular rates of the 
angular elements to be used in the present, novel formulation can be estimated very accurately 
by numerically fitting the estimated orbital ephemeris. 
Clearly, since they are based on observational data, these rates include the effects 
of \emph{all} acting secular perturbations. When applying the present formulation to
evaluate the effects, for instance, of neglected sources of perturbations along the given orbit,
it is therefore necessary to exclude all terms generating secular perturbations from the
forcing acceleration $\bmath{g}\left( \bmath{r},\bmath{\dot{r}},t\right)$ appearing 
on the right-hand side of equation (\ref{Encke EoM 2}).
In this particular application, if Encke's approach is adopted, no rectification is then needed,
since the formulation is guaranteed not to generate any secular drift between the perturbed and 
the reference orbits. Applications of this novel formulation to the case of tidal perturbations 
will be the subject of a future contribution.

\section*{Acknowledgments}

We gratefully acknowledge the support of the GOCE Italy project funded by the Italian Space Agency (ASI).

\appendix

\section{The operators $\mathcal{D}_{f}$ and $\mathcal{D}_{f}^2$ }\label{matrix_manipulation}

The operators $\mathcal{D}_{f}$ and $\mathcal{D}_{f}^{2}$ were defined in Section \ref%
{KB_reformulation} in terms of the matrices $\mathbfss{V},$ $\mathbfss{H},$ $%
\mathbfss{K}$ and $\mathbfss{N.}$ Comparison of equations (\ref{first_der_R_Mat}%
) and (\ref{sec_der_R_Mat}) with equations (\ref{D_f}) and (\ref{D2_f})
makes it clear that, since $\mathbfss{R}$ is orthogonal, these matrices are
defined as%
\begin{eqnarray}
\mathbfss{V} &\mathbfss{=}&\frac{d\mathbfss{D}}{d\Omega }\mathbfss{CBR}^{T},
\label{V_op} \\
\mathbfss{H} &=&\mathbfss{DC}\frac{d\mathbfss{B}}{du}\mathbfss{R}^{T},  \label{H_op}
\\
\mathbfss{K} &=&\frac{d^{2}\mathbfss{D}}{d\Omega ^{2}}\mathbfss{CBR}^{T},
\label{K_op} \\
\mathbfss{N} &=&\frac{d\mathbfss{D}}{d\Omega }\mathbfss{C}\frac{d\mathbfss{B}}{du}%
\mathbfss{R}^{T}.  \label{N_op}
\end{eqnarray}%
Now since $\mathbfss{R}^{T}=\mathbfss{B}^{T}\mathbfss{C}^{T}\mathbfss{D}^{T},$ it is
straightforward to show that 
\begin{eqnarray}
\mathbfss{V} &=&\frac{d\mathbfss{D}}{d\Omega }\mathbfss{D}^{T}=\left( 
\begin{array}{ccc}
0 & -1 & 0 \\ 
1 & 0 & 0 \\ 
0 & 0 & 0%
\end{array}%
\right) ,  \label{V_Mat} \\
\mathbfss{H} &=&\mathbfss{DC}\frac{d\mathbfss{B}}{du}\mathbfss{B}^{T}\mathbfss{C}^{T}%
\mathbfss{D}^{T}=\left( 
\begin{array}{ccc}
0 & -\cos i & -\sin i\cos \Omega \\ 
\cos i & 0 & -\sin i\sin \Omega \\ 
\sin i\cos \Omega & \sin i\sin \Omega & 0%
\end{array}%
\right) ,  \label{H_Mat} \\
\mathbfss{K} &\mathbfss{=}&\frac{d^{2}\mathbfss{D}}{d\Omega ^{2}}\mathbfss{D}%
^{T}=\left( 
\begin{array}{ccc}
-1 & 0 & 0 \\ 
0 & -1 & 0 \\ 
0 & 0 & 0%
\end{array}%
\right) ,  \label{K_mat} \\
\mathbfss{N} &\mathbfss{=}&\frac{d\mathbfss{D}}{d\Omega }\mathbfss{C}\frac{d\mathbfss{B%
}}{du}\mathbfss{B}^{T}\mathbfss{C}^{T}\mathbfss{D}^{T}=\left( 
\begin{array}{ccc}
-\cos i & 0 & \sin i\sin \Omega \\ 
0 & -\cos i & -\sin i\cos \Omega \\ 
0 & 0 & 0%
\end{array}%
\right) ,  \label{N_mat}
\end{eqnarray}%
where we have repeatedly used the fact that%
\begin{equation}
\frac{d\mathbfss{B}}{du}\mathbfss{B}^{T}=\left( 
\begin{array}{ccc}
0 & -1 & 0 \\ 
1 & 0 & 0 \\ 
0 & 0 & 0%
\end{array}%
\right) .  \label{dB/du Bt}
\end{equation}%
This equation, like (\ref{V_Mat}), shows a familiar property of rotation
matrices (cf. \cite{Pars}).

It remains to be shown that the action of the last operator $\mathbfss{DCB}%
_{uu}$ on the right hand side of equation (\ref{sec_der_R_Mat}) on $%
\bmath{r}_{rtn}$ is equivalent to multiplication by $-\mathbfss{R}.$ This
follows immediately once it is realized that 
\begin{equation}
\frac{d^{2}\mathbfss{B}}{du^{2}}=\mathbfss{S}-\mathbfss{B},
\end{equation}%
with%
\begin{equation}
\mathbfss{S}=\left( 
\begin{array}{ccc}
0 & 0 & 0 \\ 
0 & 0 & 0 \\ 
0 & 0 & 1%
\end{array}%
\right) .
\end{equation}%
In fact, since $\mathbfss{S}\bmath{r}_{rtn}=\mathbfss{0,}$ we have%
\begin{equation}
\mathbfss{DC}\frac{d^{2}\mathbfss{B}}{du^{2}}\bmath{r}_{rtn}=\mathbfss{DCS}%
\bmath{r}_{rtn}-\mathbfss{R}\bmath{r}_{rtn}=-\mathbfss{R}\bmath{r}%
_{rtn}=-\bmath{r}.
\end{equation}%
The full operator, analogous to $\mathbfss{V},$ $\mathbfss{H,}$ etc., of course
is defined as%
\begin{equation}
\mathbfss{DC}\frac{d^{2}\mathbfss{B}}{du^{2}}\mathbfss{R}^{T}=\mathbfss{DCSR}^{T}-%
\mathbfss{I},  \label{DCBuuRt}
\end{equation}%
but for the present purposes it is expedient to use its restriction to
vectors having a null third component, which justifies our definition of the
operator in equation (\ref{D2_f}).

\section{The operators $\mathcal{D}_{t}$ and $\mathcal{D}_{t}^2$  }\label{matrix_manipulation_MA}

For the first derivatives of rotation matrix $\mathbfss{R}$ (\ref{R_Mat}) with
respect to time we have%
\begin{equation}
\frac{d\mathbfss{R}}{dt}=\frac{d\mathbfss{D}}{dt}\mathbfss{CB}+\mathbfss{DC}\frac{d%
\mathbfss{B}}{dt}.  \label{dR_dt}
\end{equation}%
The first term to the left hand side can be rewritten as%
\begin{equation}
\frac{d\mathbfss{D}}{dt}\mathbfss{CB}=\dot{\Omega}\frac{d\mathbfss{D}}{d\Omega }%
\mathbfss{D}^{T}\mathbfss{DCB}=\dot{\Omega}\mathbfss{VR},
\end{equation}%
where $\mathbfss{V}$ is the matrix already introduced in Eq. (\ref{V_Mat}).

The second term of (\ref{dR_dt}) can be reformulated as%
\begin{equation}
\mathbfss{DC}\frac{d\mathbfss{B}}{dt}=\dot{u}\mathbfss{DC}\frac{d\mathbfss{B}}{du}=(%
\dot{\omega}+\dot{f})\mathbfss{HR},
\end{equation}%
where $\mathbfss{H}$ is given by Eq. (\ref{H_Mat}). Collecting terms we have
that the first derivatives of $\mathbfss{R}$ is%
\begin{equation}
\frac{d\mathbfss{R}}{dt}=\left[ (\dot{\omega}+\dot{f})\mathbfss{H}+\dot{\Omega}%
\mathbfss{V}\right] \mathbfss{R}.  \label{dR_dt_final}
\end{equation}

The second derivatives of $\mathbfss{R}$ can be computed from previous
equation that is 
\begin{equation}
\frac{d^{2}\mathbfss{R}}{dt^{2}}=\frac{d}{dt}\left[ (\dot{\omega}+\dot{f})%
\mathbfss{H}+\dot{\Omega}\mathbfss{V}\right] \mathbfss{R+}\left[ (\dot{\omega}+%
\dot{f})\mathbfss{H}+\dot{\Omega}\mathbfss{V}\right] ^{2}\mathbfss{R}.
\label{d2R_dt2}
\end{equation}%
The first term of right hand side can be computed as%
\begin{equation}
\frac{d}{dt}\left[ (\dot{\omega}+\dot{f})\mathbfss{H}+\dot{\Omega}\mathbfss{V}%
\right] =\ddot{f}\mathbfss{H}+\dot{\Omega}(\dot{\omega}+\dot{f})\mathbfss{T},
\end{equation}%
where we have substituted for%
\begin{equation}
{\dot\mathbfss{H}}=\dot{\Omega}\mathbfss{T}\,,
\end{equation}%
with%
\begin{equation}
\mathbfss{T}=\left( 
\begin{array}{ccc}
0 & 0 & \sin i\sin \Omega \\ 
0 & 0 & -\sin i\cos \Omega \\ 
-\sin i\sin \Omega & \sin i\cos \Omega & 0%
\end{array}%
\right) .  \label{T_Mat}
\end{equation}

Substituting and collecting terms in equation (\ref{d2R_dt2}) yields 
\begin{equation}
\frac{d^{2}\mathbfss{R}}{dt^{2}}=\left[ (\dot{\omega}+\dot{f})^{2}\mathbfss{H}%
^{2}+2\dot{\Omega}(\dot{\omega}+\dot{f})\mathbfss{N}+\dot{\Omega}^{2}\mathbfss{V}%
^{2}+\ddot{f}\,\mathbfss{H}\right] \mathbfss{R},  \label{d2R_dt2_final}
\end{equation}%
where we have used%
\begin{equation}
\mathbfss{Q}=\mathbfss{HV}+\mathbfss{VH}=\left( 
\begin{array}{ccc}
-2\cos i & 0 & \sin i\sin \Omega \\ 
0 & -2\cos i & -\sin i\cos \Omega \\ 
\sin i\sin \Omega & -\sin i\cos \Omega & 0%
\end{array}%
\right) ,  \label{Q_Mat}
\end{equation}%
and%
\begin{equation}
\mathbfss{Q}+\mathbfss{T}=2\mathbfss{N},  \label{Q_Mat+T_Mat = N/2}
\end{equation}%
with%
\begin{equation}
\mathbfss{N}=\left( 
\begin{array}{ccc}
-\cos i & 0 & \sin i\sin \Omega \\ 
0 & -\cos i & -\sin i\cos \Omega \\ 
0 & 0 & 0%
\end{array}%
\right) .  \label{N_mtx}
\end{equation}

It is now possible to define the operators $\mathcal{D}_{t}$ and $\mathcal{D}%
_{t}^{2}$ as%
\begin{eqnarray}
\mathcal{D}_{t} &=&(\dot{\omega}+\dot{f})\mathbfss{H}+\dot{\Omega}\mathbfss{V},
\label{D_t Appendix} \\
\mathcal{D}_{t}^{2} &=&(\dot{\omega}+\dot{f})^{2}\mathbfss{H}^{2}+2\dot{\Omega}%
(\dot{\omega}+\dot{f})\mathbfss{N}+\dot{\Omega}^{2}\mathbfss{K}+\ddot{f}\,%
\mathbfss{H},  \label{D2_t Appendix}
\end{eqnarray}%
where we have used the identity $\,\allowbreak $%
\begin{equation}
\mathbfss{V}^{2}=\mathbfss{K}.  \label{V2=K}
\end{equation}%
Note that these operators are intended for application to a position vector
only. A further relationship is useful in connection with the operator $%
\mathbfss{H}^{2}.$ If we insert the identity $\mathbfss{BB}^{T}$ after the
factor $\mathbfss{C}$ in equation (\ref{H_op}) and note that $\mathbfss{B}^{T}(d%
\mathbfss{B}/du)=\mathbfss{V,}$ it follows that%
\begin{equation}
\mathbfss{H}=\mathbfss{RVR}^{T}.  \label{H=RVRt}
\end{equation}%
Then, taking the square of both sides and multiplying by $\mathbfss{R}$ on the
right we find%
\begin{equation}
\mathbfss{H}^{2}\mathbfss{R}=\mathbfss{RV}^{2}\mathbfss{R}^{T}\mathbfss{R}=\mathbfss{RK}.
\label{H2R}
\end{equation}

\bsp

\label{lastpage}

\end{document}